\documentclass[9pt,conference]{IEEEtran}
\IEEEoverridecommandlockouts

\usepackage{microtype}
\usepackage{xurl}
\usepackage{cite}
\usepackage{bm,amsmath,amssymb,amsfonts}
\usepackage{algorithmic}
\usepackage{graphicx}
\usepackage{textcomp}
\usepackage{xcolor}
\usepackage[normalem]{ulem}
\useunder{\uline}{\ul}{}
\def\BibTeX{{\rm B\kern-.05em{\sc i\kern-.025em b}\kern-.08em
    T\kern-.1667em\lower.7ex\hbox{E}\kern-.125emX}}
\begin{document}

\title{Compositional Audio Representation Learning
}


\author{
\IEEEauthorblockN{
Sripathi Sridhar,
Mark Cartwright
}\\
\IEEEauthorblockA{
    Sound Interaction and Computing (SInC) Lab, New Jersey Institute of Technology, Newark, NJ, USA \\
    \{ss645, mark.cartwright\}@njit.edu
    }
}

\maketitle

\begin{abstract}
Human auditory perception is compositional in nature --- we identify auditory streams from auditory scenes with multiple sound events. However, such auditory scenes are typically represented using clip-level representations that do not disentangle the constituent sound sources. In this work, we learn source-centric audio representations where each sound source is represented using a distinct, disentangled source embedding in the audio representation. We propose two novel approaches to learning source-centric audio representations: a supervised model guided by classification and an unsupervised model guided by feature reconstruction, both of which outperform the baselines. We thoroughly evaluate the design choices of both approaches using an audio classification task. We find that supervision is beneficial to learn source-centric representations, and that reconstructing audio features is more useful than reconstructing spectrograms to learn unsupervised source-centric representations. Leveraging source-centric models can help unlock the potential of greater interpretability and more flexible decoding in machine listening.
\end{abstract}


\section{Introduction}
\label{sec:intro}

Human auditory perception is compositional in nature---we tend to understand our environments by grouping and segregating auditory ‘streams’ from the available information in the scene \cite{bregman_auditory_1994}. While auditory scenes often contain overlapping sound sources, audio representation models typically encode the scene using a clip-level representation that does not delineate between constituent sound sources, making it challenging to do source-level inference. This limits machine listening's potential in applications such as individual ID \cite{knight2024individual}, sound localization, and tracking sound sources at different levels of granularity towards a compositional representation of auditory scenes, as is now commonplace in computer vision \cite{chen_compositional_2024, webb_systematic_2023}.

The machine listening task of source separation aims to estimate component sources from an audio mixture, typically by generating source-specific masks on the input mixture audio in a shallow latent space \cite{zeghidour_wavesplit_2021, StollerED18-0} or the time-frequency domain \cite{wisdom_whats_2021, petermann_hyperbolic_2023, liu_separate_2023}. However, source separation models do not aim to learn semantic representations for use in downstream audio understanding tasks such as audio captioning, classification, or retrieval. Furthermore, solving universal source separation is a challenging endeavor that may not be necessary to advance machine listening toward compositional representations for source inference. To develop models that better align with the compositional nature of human perception, we propose to directly learn compositional audio representations, where the model encodes the semantic content of each sound source in the scene with a corresponding source embedding, and each source embedding can be flexibly decoded depending on the downstream task, e.g., tracking an individual animal in a dense forest soundscape. Each source can be seen as an auditory ‘object’ under this framework.

\textit{Object-centric learning} (OCL) in computer vision, also known as \textit{compositional scene representation learning}, aims to decompose a visual scene into independent representations of its composite visual objects \cite{yuan_compositional_2023}. Visual OCL has been successfully applied to tasks such as scene segmentation, object discovery, and object property prediction \cite{locatello_object-centric_2020, carion_end--end_2020, seitzerbridging, singh_simple_2022, brady_provably_2023, engelcke_genesis-v2_2021}. In OCL, a slot attention module is typically used to transform an image-level representation to an object-centric one where each object is encoded in a ‘slot’ \cite{ locatello_object-centric_2020}. In the unsupervised setup, such representations were initially learned by reconstructing the input image pixels \cite{locatello_object-centric_2020, singh_simple_2022, brady_provably_2023, engelcke_genesis-v2_2021, carion_end--end_2020} and more recently by reconstructing semantic features \cite{seitzerbridging, singh2022illiterate}. Seitzer et al. showed that feature reconstruction may be more useful than image reconstruction for object-centric models to generalize to real-world image datasets.
\begin{figure}
    \centering
    \includegraphics[width=\columnwidth]{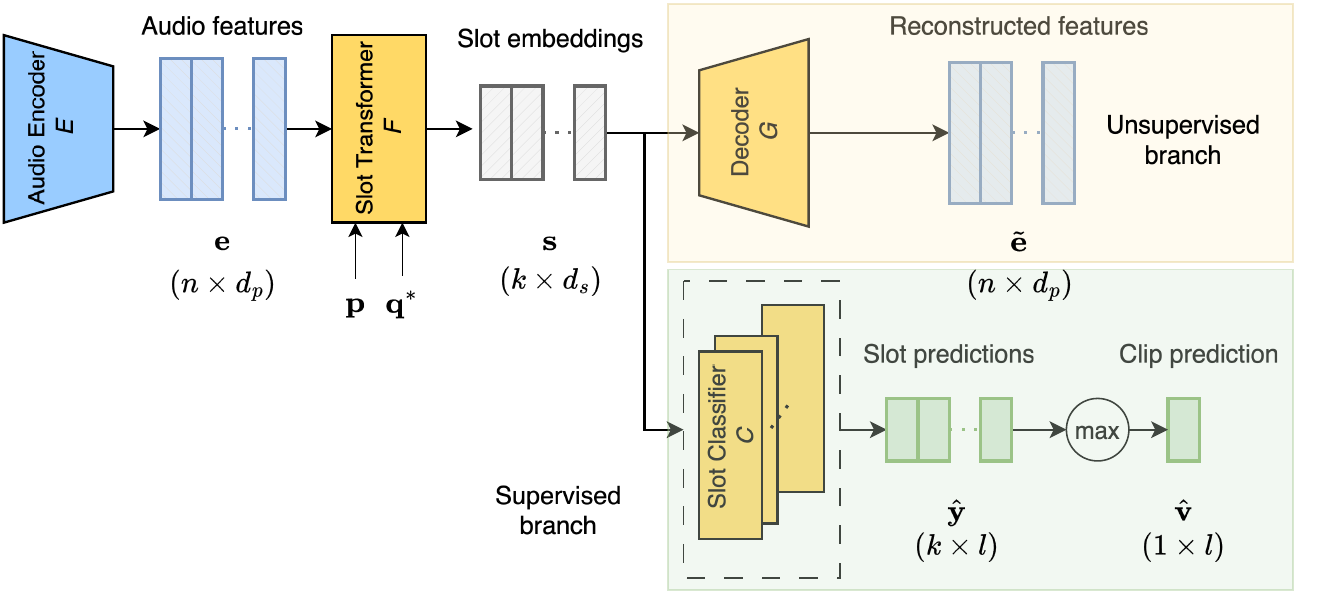}
    \caption{Proposed compositional audio representation learning (CARL) model consisting of a frozen audio encoder, slot transformer, reconstruction decoder, and classifier. The modules in yellow are trained jointly using a combined weighted loss function, including a reconstruction objective on the unsupervised branch, and a permutation-invariant cross-entropy loss on the supervised branch.}
    \label{fig:supervised-carl}
\end{figure}

OCL has recently begun to gain interest in machine listening as well. Reddy et al. \cite{reddy_audioslots_2023} introduced a strongly supervised source separation model that separately encodes each sound source in the auditory scene, then decodes to source-level spectrograms. While this model is evaluated on a two-speaker separation task, the representations are not evaluated on downstream audio understanding tasks.
Gha et al. \cite{gha_unsupervised_2023} use slot attention for musical note detection, where the goal is to predict spectrograms of all the notes present in a musical chord. They find their unsupervised model outperforms supervised classifiers and autoencoder models that do not learn object-centric representations. The model treats each note in a musical chord as an object and is evaluated on a `note property prediction’ task on synthetic data generated with virtual instruments. While promising, it is unclear if their approach would generalize to real-world music or environmental sound with more diverse time-frequency structure. Thus, object-centric learning, or more aptly named \textit{source-centric representation learning}, remains an open challenge in machine listening.

In this work, we aim to advance the development of source-centric representation learning. To do so, we:
\begin{enumerate}
    \item Propose source-centric audio representation learning for environmental sound---a novel task
    \item Investigate the design space of these models in terms of choices such as the reconstruction target.
    \item Evaluate unsupervised and supervised source-centric audio representations on multi-label audio classification.
\end{enumerate}


\section{Compositional Audio Representation Learning}

We provide an overview of the source-centric representation learning, or \textit{compositional audio representation learning (CARL)} framework shown in Figure~\ref{fig:supervised-carl}. Given an input audio spectrogram $\mathbf{x} \in \mathbb{R}^{t\times f}$, the audio encoder $E$  generates audio features $\mathbf{e} = E(\mathbf{x}) \in \mathbb{R}^{n \times d_p}$, consisting of patch embeddings where $n$ is the number of spectro-temporal patches, and $d_p$ is the patch embedding dimensionality. The slot transformer $F$, conditioned on a fixed 2-D time-frequency sinusoidal positional encoding $\mathbf{p}$ and a learned initial query $\mathbf{q}^*$, transforms these features into slot embeddings $\mathbf{s} = F(\mathbf{e} | \mathbf{p}, \mathbf{q}^*) \in \mathbb{R}^{k \times d_s}$, where  $k$ is the number of slots and $d_s$ is the slot embedding dimensionality. 
The decoder $G$ reconstructs the target from $\mathbf{s}$, and the target can be set to either $\mathbf{e}$ or $\mathbf{x}$. The classifier $C$ generates a multi-label prediction $\hat{\mathbf{v}} = C(\mathbf{s}) \in \mathbb{R}^{1 \times l}$, where $l$ is the vocabulary size, by aggregating $k$ slot-level predictions.

\subsection{Audio encoder}
We use the AudioMAE audio encoder trained with self-supervision that has been shown to perform well on a variety of downstream tasks \cite{huang_masked_2022}. 
Moreover, AudioMAE features are patch embeddings that retain time-frequency spatial information, although at a lower resolution than typical spectrograms, which could be useful to delineate overlapping sound events in the audio clip. 

\subsection{Slot transformer} 
The slot transformer module decomposes the audio features into slot embeddings. In the OCL approach proposed by Locatello et al. \cite{locatello_object-centric_2020}, a \textit{slot attention} module 
is implemented using a GRU, where each slot competes to represent objects in the input image, and the slot embeddings are iteratively refined over multiple time steps. Recent studies have used this slot attention module \cite{locatello_object-centric_2020} on object discovery \cite{seitzerbridging} and musical note detection tasks \cite{gha_unsupervised_2023}.
Prior work also evaluated a transformer-based attention module on object discovery \cite{carion_end--end_2020} and audio source separation \cite{reddy_audioslots_2023} suggesting its usefulness for CARL. Wu et al. \cite{wu_inverted-attention_2023} showed that a transformer with inverted attention is conceptually similar to the original slot attention module, and achieves similar object detection performance on various synthetic image datasets. We opt for a \textit{slot transformer} in this work due its greater modeling capacity over a GRU with shared-weights layers in the original slot attention module \cite{locatello_object-centric_2020}. Specifically, we use an inverted-attention pre-normalization transformer decoder with $m+2 = 6$ slots, where $m = 4$ is the maximum polyphony of the dataset. The additional slots may be useful to encode noise and semantic information that doesn't belong to any of the desired sound event categories \cite{gha_unsupervised_2023}. 

\subsection{Reconstruction decoder}
The reconstruction decoder reconstructs a target consisting of either the input audio features $\mathbf{e}$ or input spectrogram $\mathbf{x}$. We explore both spatial broadcast and autoregressive decoders in this work. 

The spatial broadcast decoder broadcasts the slot embeddings to the target 2-D dimensions, e.g., for patch embeddings or a spectrogram. It estimates either $\tilde{\mathbf{e}}_{i} \in \mathbb{R}^{n \times d_p}$ or $\tilde{\mathbf{x}}_i \in \mathbb{R}^{t\times f}$ for slot $i$, and a corresponding alpha mask $\boldsymbol{\alpha}_i \in \mathbb{R}^{n \times 1}$ or $\boldsymbol{\alpha}_i \in \mathbb{R}^{t \times f}$, for feature reconstruction and spectrogram reconstruction, respectively. Alpha mask, originally proposed by Locatello et al. \cite{locatello_object-centric_2020}, is a single channel mask used after softmax normalization as weights to composite reconstructed slot-level features to clip-level features $\tilde{\mathbf{e}} = \sum_{i=1}^{k}\Bar{\boldsymbol{\alpha}}_i \odot \tilde{\mathbf{e}}_{i}$, where $k$ is the number of slots and $\Bar{\boldsymbol{\alpha}}_i$ is the alpha mask after softmax normalization across slots. 
For spectrogram reconstruction, we use a CNN spatial broadcast decoder \cite{locatello_object-centric_2020, reddy_audioslots_2023}, similar to the one proposed by Gha et al. \cite{gha_unsupervised_2023}, and for AudioMAE feature reconstruction, we use an MLP spatial broadcast decoder based on the one used by Seitzer et al \cite{seitzerbridging}. For spectrogram reconstruction, we first convert the log-domain reconstructions to the linear domain before summing, then convert back to compute the loss. The autoregressive transformer decoder was originally explored for image generation \cite{singh2022illiterate} and then used for object discovery by Seitzer et al. \cite{seitzerbridging}. The transformer decoder autoregressively reconstructs features jointly across all slots, conditioned on the set of previously generated audio features. For AudioMAE feature reconstruction, we also evaluate a transformer decoder with four pre-normalization layers and compare it to the MLP spatial broadcast decoder.

\subsection{Classifier}
For providing additional guidance for source-centric learning, we use a linear shared-weights classifier with one head per slot (see Figure~\ref{fig:supervised-carl}) to generate slot-level predictions during training and inference. 
 Each slot embedding is independently transformed by its corresponding classifier head to generate the slot prediction. 

\subsection{Loss terms}
We use a weighted combined loss that includes a classifier loss, decoder reconstruction loss, as well as two additional penalty terms. The decoder is trained using a mean squared error reconstruction loss. 
The classifier is trained using a permutation-invariant cross-entropy loss, where the best match between the slot-level predictions and the ground-truth labels is used to compute the clip-level loss.
We introduce a \textit{sparsity penalty}, computing the mean $L_1$ norm on $k$ slot-level classifier predictions $\hat{\mathbf{y}}_{i}$ as $L_s = \frac{1}{k}\sum_{i=1}^{k}\Vert \hat{\mathbf{y}}_{i} \Vert_1$ to encourage each slot to encode a single source.
We also introduce a \textit{disjointedness penalty}, to discourage multiple slots from encoding the same semantic content. We compute similarity $\phi_{i,j}$ between all pairs of slot embeddings ($\mathbf{s}_i, \mathbf{s}_j$): $\phi_{i,j} = \operatorname{max}(0, \frac{\mathbf{s}_i.\mathbf{s}_j}{\Vert{\mathbf{s}_i}\Vert\Vert{\mathbf{s}_j}\Vert})
        $ and then compute the clip-level sum: $
\phi = \sum_{i=1}^{k-1}\sum_{j=i+1}^{k}\phi_{i,j}.
$
When labels are available, we only compute disjointedness for slots which are matched with a ground-truth, so as not to penalize the similarity of two `inactive' slots. We compute disjointedness on a linear projection of slot embeddings. 

\section{Evaluation}
\subsection{Data}
We evaluate the models on OST\footnote{doi.org/10.5281/zenodo.13755902}, a synthetic multi-label audio classification dataset of 500k 1-second samples generated from FSD-50k with Scaper \cite{salamon_scaper_2017}, presented in previous work by the authors \cite{Sridhar2023}. We refer the reader to this work for further details. We also introduce a smaller vocabulary version of OST with 10 classes, called OST-Tiny, to investigate the effect of vocabulary size with a similar data distribution. For compatibility with AudioMAE, we resample all audio files to 16kHz and compute mel-spectrograms using 128 mel bins, window size 25 ms, and hop size 10 ms.

We pick these datasets for two reasons: (1) They provide a controlled setting to evaluate CARL. (2) The embeddings can be evaluated on unseen classes, simulating a more realistic inference scenario.

\subsection{Training}
\label{sec:training}
We tune hyperparameters using Optuna \cite{akiba_optuna_2019} and set the slot dimension and projection dimension to 512 and 16 respectively, for OST. We use slot dimension 128 for OST-Tiny, and did not find the projection layer to be useful. For the supervised model, we set the loss weights to 100 for disjointedness, 0.1 for reconstruction and sparsity, and 1 for cross-entropy.

The unsupervised models are trained using a two-stage approach. First, we train the slot transformer and decoder using a weighted combination of reconstruction and disjointedness losses. Then, we discard the decoder and train a linear classifier on top of the frozen slot embeddings using a cross-entropy loss. Since label supervision is unavailable, we compute the cosine similarity between all pairs of slot embeddings. After a separate round of hyperparameter tuning on OST, we use a disjointedness weight of 0.01 and a reconstruction weight of 1 for the unsupervised CARL models.

\subsection{Evaluation metrics}
We evaluate the source-centric representation capabilities of our proposed models on a multi-label classification task. We employ a `max-per-slot then aggregate' strategy to compute the clip-level multi-label prediction, to better capture contributions from each slot. The maximum softmax normalized prediction is computed for each slot, and the remaining class predictions are set to zero. A $\operatorname{max}$ operator is then used to aggregate class-level predictions across $k$ slots to form a multi-label clip-level prediction. We evaluate classification performance with mAP.

As a secondary metric, we compute the supervised silhouette score \cite{rousseeuw_silhouettes_1987} on the slot embeddings to estimate semantic cluster quality. Since slot embeddings are unordered, we compute the best match between slot-level classifier predictions and the ground truth, and permute the slot embeddings to reflect that ordering. Each slot embedding is then independently evaluated against its best-matched ground truth.

\subsection{Baselines}
We compare against a model that uses universal source separation (USS) to validate the effectiveness of CARL. 
We train a MixIT TDCN++ model \cite{wisdom_unsupervised_2020} without labels on OST. We then encode the source estimates with AudioMAE and use them to train a shared-weights linear classifier with one head per source estimate on OST using permutation-invariant cross-entropy loss. 

We also evaluate a mixture (i.e., unseparated) AudioMAE baseline to simulate an ineffective slot transformer. AudioMAE embeddings are copied to each slot and input to a shared-weights linear classifier with one head per slot, trained on OST. Both baselines use the same training hyperparameters as the unsupervised CARL classifier. 

\subsection{Experiments}
\label{sec:experiments}
We run a series of experiments to evaluate supervised and unsupervised CARL and to investigate the design space of these models. Unless mentioned otherwise, we evaluate models only on seen classes from the evaluation set of OST and OST-Tiny.

\smallskip
\noindent\textbf{Exp 1:} Supervised CARL design choices. We evaluate the full supervised CARL model seen in Figure~\ref{fig:supervised-carl} on OST and OST-Tiny and compare three reconstruction decoder types: a transformer feature decoder and an MLP feature decoder with and without alpha masks. We first train these models with the combined loss function described in Section~\ref{sec:training} and evaluate them using mAP and silhouette score.

We subsequently run loss ablations for supervised CARL, evaluating the relative importance of reconstruction, disjointedness and sparsity losses through removal. We train the models on OST and evaluate performance using mAP and silhouette score.

\smallskip
\noindent\textbf{Exp 2:} Unsupervised CARL design choices. We investigate the effect of target type and decoder type on unsupervised CARL. We compare feature reconstruction and spectrogram reconstruction as target types, as well as an autoregressive transformer against a spatial broadcast decoder as decoder types. Specifically, we evaluate five decoders and two target types: a transformer feature decoder, an MLP feature decoder with and without alpha, and a CNN spectrogram decoder with and without alpha. All models are trained on OST using the protocol described in Section~\ref{sec:training}, and evaluated using mAP.

We further explore the design space for unsupervised CARL by running ablations on 1) inverted attention by comparing to a standard transformer, 2) the disjointedness loss term by removal, and 3) slot embedding dimensionality by varying the size. We train the models on OST and evaluate performance using mAP.

\smallskip
\noindent\textbf{Exp 3:} Comparison to baselines. To validate the effectiveness of the proposed models, we compare CARL against the USS and AudioMAE baselines. Baselines are trained on OST and evaluated using mAP.

\smallskip
\noindent\textbf{Exp 4:} Generalization. Lastly, we evaluate the generalization of supervised CARL models to unseen classes to simulate inference in more realistic open-set machine listening scenarios. First we train the models on OST training data as described in Section~\ref{sec:training}. Then we discard the classifier, freeze the embeddings, and train a new shared-weights linear classifier on the evaluation set of OST, which contains the 53 seen classes from the training set as well as 34 unseen classes. We then evaluate model performance using mAP on a held out portion of the evaluation set. 

\section{Results}
\subsection{Design choices for supervised CARL}
\tabcolsep=0.13cm
\begin{table}[]
\centering
\caption{\normalfont{Results on supervised CARL with the combined loss including reconstruction, disjointedness and sparsity. 
Silhouette score (ss) is computed on the slot embeddings using the best match between the predictions and ground truth.}}
\label{tab:supervised-big-and-tiny}
\begin{tabular}{cclccccc}
\hline
\textbf{Supervised} & \textbf{Decoder} & \multicolumn{1}{c}{\textbf{Target}} & \textbf{Alpha} & \multicolumn{2}{c}{\textbf{OST}} & \multicolumn{2}{c}{\textbf{OST-Tiny}} \\ \hline
 & \textbf{} &  & \textbf{} & \textbf{mAP} & \textbf{ss} & \textbf{mAP} & \textbf{ss} \\ \hline
Y & MLP & Features & Y & 0.677 & 0.227 & 0.814 & 0.393 \\
Y & MLP & Features & N & 0.685 & 0.256 & 0.797 & 0.392 \\
Y & Transformer & Features & -- & \textbf{0.698} & \textbf{0.282} & \textbf{0.849} & \textbf{0.476} \\ \hline
\end{tabular}
\end{table}

 As seen in Table~\ref{tab:supervised-big-and-tiny}, we find that the supervised CARL model with a transformer decoder performs best on both OST and OST-Tiny, both in terms of classification performance and clustering quality. We also find that the MLP decoder without alpha mask performs better than the model with the alpha mask. 

\begin{table}[]
\centering
\caption{\normalfont{Loss ablations on the supervised transformer feature decoder model trained with label supervision, evaluated on OST. ss refers to the silhouette score.}}
\label{tab:loss-ablations}
\begin{tabular}{cccccc}
\hline
\textbf{Supervised} & \textbf{Decoder} & \textbf{Target} & \textbf{Ablation} & \textbf{mAP} & \textbf{ss} \\ \hline
Y & Transformer & Features & --- & \textbf{0.698} & \textbf{0.282} \\ \hline
Y & Transformer & Features & Reconstruction & 0.684 & 0.279 \\
Y & Transformer & Features & Disjointedness & 0.662 & 0.280 \\
Y & Transformer & Features & Sparsity & 0.629 & 0.247 \\ \hline
\end{tabular}
\end{table}

We investigate the usefulness of the loss terms for supervised CARL in Table~\ref{tab:loss-ablations}. We find that sparsity is particularly useful, both to classification performance and clustering quality. Reconstruction is the least useful, which indicates that label supervision may be the dominant learning signal. Although disjointedness is useful to classification performance, it doesn't affect clustering quality significantly. During initial experiments on OST-Tiny, we found disjointedness to be useful for clustering quality too, suggesting that the vocabulary size may be a factor.

\subsection{Design choices for unsupervised CARL}
\label{sec:targetxdecoder} 

\tabcolsep=0.12cm
\begin{table}[]
\centering
\caption{\normalfont{Effect of target and decoder for unsupervised CARL evaluated on OST, broken down by polyphony, i.e. the number of tags. 'all' refers to the full evaluation set, while 'p1', 'p2', and 'p3' refer to the examples with one, two, and three tags. USS refers to a universal source separator.}}
\label{tab:targetxdecoder}
\begin{tabular}{cccccccc}
\hline
\textbf{Sup.} & \textbf{Decoder/Model} & \textbf{Target} & \textbf{Alpha} & \multicolumn{4}{c}{\textbf{mAP by polyphony}} \\ \hline
 &  &  &  & \textbf{all} & \textbf{p1} & \textbf{p2} & \textbf{p3} \\ \hline
N & MLP & Features & Y & 0.530 & 0.580 & 0.409 & 0.330 \\
N & MLP & Features & N & \textbf{0.578} & \textbf{0.629} & \textbf{0.454} & \textbf{0.371} \\
N & Transformer & Features & -- & 0.516 & 0.569 & 0.391 & 0.324 \\ \hline
N & CNN & Spectrogram & Y & 0.428 & 0.486 & 0.292 & 0.261 \\
N & CNN & Spectrogram & N & 0.453 & 0.512 & 0.307 & 0.261 \\ \hline
N & USS & -- & -- & 0.370 & 0.394 & 0.316 & 0.323 \\
N & AudioMAE & -- & -- & 0.476 & 0.554 & 0.262 & 0.182 \\ \hline
\end{tabular}
\end{table}

We present the results in Table~\ref{tab:targetxdecoder} broken down by polyphony. We find that feature reconstruction is significantly more effective than spectrogram reconstruction for learning unsupervised source-centric representations. Moreover, the MLP feature decoder outperforms the transformer-based feature decoder. This is in line with prior results on unsupervised visual object discovery \cite{seitzerbridging}. Similar to the supervised model results in Table~\ref{tab:supervised-big-and-tiny}, models without an alpha mask perform better than those using a softmax normalized alpha mask. This validates a similar finding by Gha et al. \cite{gha_unsupervised_2023} --- they speculate that a spectrogram decoder without an alpha mask may perform better on examples with overlapping sources due to the fact that auditory objects overlap in time and frequency and may require multiple active slots per time-frequency region. However, when we break down our results by polyphony, we find that a spectrogram decoder without an alpha mask doesn't outperform one with an alpha mask on higher polyphony examples (see column p3 in Table~\ref{tab:targetxdecoder}) but does so on monophonic examples. Thus, further analysis is needed to understand why this may be the case.


We evaluate the best unsupervised CARL model from prior experiments, MLP feature decoder without alpha, in Table~\ref{tab:slot-dim-ablation}. We find a monotonic decrease in classification performance as the slot embedding dimensionality $d_s$ decreases, suggesting that higher dimensional slots may be useful to learn source-specific embeddings.

\begin{table}[]
\centering
\caption{\normalfont{Varying the slot embedding dimensionality for the best performing unsupervised model, MLP feature decoder without alpha, evaluated on OST.}}
\label{tab:slot-dim-ablation}
\begin{tabular}{cccccc}
\hline
\textbf{Supervised} & \textbf{Decoder} & \textbf{Target} & \textbf{Alpha} & \textbf{$\mathbf{d_s}$} & \textbf{mAP} \\ \hline
N & MLP & Features & N & 512 & \textbf{0.578} \\ \hline
N & MLP & Features & N & 256 & 0.523 \\
N & MLP & Features & N & 64 & 0.424 \\ \hline
\end{tabular}
\end{table}

We also investigate the choice of inverted attention in the slot transformer and the disjointedness penalty in Table~\ref{tab:ablation-unsupervised}. We see that inverted attention is useful, while disjointedness is not. This may be because disjointedness in the unsupervised model may inadvertently penalize inactive slots being similar to each other, unlike the supervised case in Table~\ref{tab:loss-ablations} where best-matching is used to ignore inactive slots.
\begin{table}[]
\centering
\caption{\normalfont{Ablations on an unsupervised MLP feature decoder, evaluated on OST. Slot transformer and decoder are trained using a reconstruction objective, then a classifier is trained on top of frozen embeddings using label supervision.}}
\label{tab:ablation-unsupervised}
\begin{tabular}{cccccc}
\hline
\textbf{Supervised} & \textbf{Decoder} & \textbf{Target} & \textbf{Alpha} & \textbf{Ablation} & \textbf{mAP} \\ \hline
N & MLP & Features & N & --- & 0.578 \\ \hline
N & MLP & Features & N & Inv. Attention & 0.562 \\
N & MLP & Features & N & Disjointedness & \textbf{0.589} \\ \hline
\end{tabular}
\end{table}

\subsection{Comparing performance against baselines}
Unsupervised CARL also yields better results compared to a USS baseline (see Table~\ref{tab:targetxdecoder}), suggesting its utility over a USS-based approach for tasks like classification. Also, the proposed feature reconstruction models perform better than an AudioMAE baseline, indicating that the slot transformer learns useful transformations on clip-level features. 

We note that although the AudioMAE baseline has an overall performance comparable to the feature reconstruction models, this is mainly limited to polyphony 1 examples, and it has a sharp drop-off for higher polyphony examples. On the other hand, source-centric representations do much better at higher polyphony, highlighting their utility over clip-level representations. We also note that the AudioMAE baseline outperforms the spectrogram decoder on polyphony 1, indicating that spectrogram reconstruction may not be a strong enough signal to learn source-centric representations.

\subsection{Generalization to unseen classes}
\label{sec:unseen}
We present evaluation results on seen and unseen classes in Table~\ref{tab:unseen}, which indicate that supervised CARL embeddings can generalize to unseen classes. Overall, we note similar trends as seen in Section~\ref{sec:targetxdecoder}---MLP without alpha performs better than with alpha, both of which outperform the transformer decoder with the combined loss. Surprisingly however, models without disjointedness and sparsity significantly outperform the model with the combined loss. This suggests that the loss function for supervised CARL should be selected based on need to generalize to unseen classes.
\tabcolsep=0.14cm
\begin{table}[]
\centering
\caption{\normalfont{Generalization to unseen classes of the supervised CARL models. All models are trained on OST (see Section \ref{sec:unseen} for more details) using a combined loss except for the models with ablations.}}
\label{tab:unseen}
\begin{tabular}{cccccc}
\hline
\textbf{Supervised} & \textbf{Decoder} & \textbf{Target} & \textbf{Alpha} & \textbf{Ablation} & \textbf{mAP} \\ \hline
Y & Transformer & Features & -- & -- & 0.571 \\
Y & Transformer & Features & -- & Reconstruction & 0.571 \\
Y & Transformer & Features & -- & Disjointedness & \textbf{0.697} \\
Y & Transformer & Features & -- & Sparsity & 0.630 \\
Y & MLP & Features & Y & -- & 0.635 \\
Y & MLP & Features & N & -- & {\ul 0.644} \\ \hline
\end{tabular}
\end{table}

\section{Discussion and Conclusion}
The results of our experiments demonstrate the utility of compositional audio representation learning and source-centric embeddings for environmental sound, and they provide guidance on the design choices of these novel models. However, there is a notable gap between the performance of the supervised and unsupervised models. This gap indicates the need for additional signal to guide the model's notion of a source and regularize the semantic source-centric embedding space. In future work, we will explore different forms of supervision to better guide CARL, e.g., source counts, partial labels, natural language, and self-supervision, and evaluate on real-world data. 

In conclusion, we evaluated supervised and unsupervised CARL models and found that they learn useful source-centric representations on synthetic data, outperforming the baselines by a significant margin. We investigated the design space of these models, finding that the reconstruction decoder target is a critical choice for unsupervised CARL, with feature reconstruction significantly outperforming spectrogram reconstruction. This work provides a solid foundation for future work in source-centric learning which we hope will enable new classes of machine listening tasks focusing on source-level inference and unlock new applications of machine listening. 
\bibliographystyle{IEEEtran}
\bibliography{IEEEabrv,references,ref2}

\begin{thebibliography}{10}
\providecommand{\url}[1]{#1}
\csname url@samestyle\endcsname
\providecommand{\newblock}{\relax}
\providecommand{\bibinfo}[2]{#2}
\providecommand{\BIBentrySTDinterwordspacing}{\spaceskip=0pt\relax}
\providecommand{\BIBentryALTinterwordstretchfactor}{4}
\providecommand{\BIBentryALTinterwordspacing}{\spaceskip=\fontdimen2\font plus
\BIBentryALTinterwordstretchfactor\fontdimen3\font minus \fontdimen4\font\relax}
\providecommand{\BIBforeignlanguage}[2]{{%
\expandafter\ifx\csname l@#1\endcsname\relax
\typeout{** WARNING: IEEEtran.bst: No hyphenation pattern has been}%
\typeout{** loaded for the language `#1'. Using the pattern for}%
\typeout{** the default language instead.}%
\else
\language=\csname l@#1\endcsname
\fi
#2}}
\providecommand{\BIBdecl}{\relax}
\BIBdecl

\bibitem{bregman_auditory_1994}
A.~S. Bregman, \emph{\BIBforeignlanguage{en}{Auditory {Scene} {Analysis}: {The} {Perceptual} {Organization} of {Sound}}}.\hskip 1em plus 0.5em minus 0.4em\relax MIT Press, Sep. 1994, google-Books-ID: jI8muSpAC5AC.

\bibitem{knight2024individual}
E.~Knight, T.~Rhinehart, D.~R. de~Zwaan, M.~J. Weldy, M.~Cartwright, S.~H. Hawley, J.~L. Larkin, D.~Lesmeister, E.~Bayne, and J.~Kitzes, ``Individual identification in acoustic recordings,'' \emph{Trends in Ecology \& Evolution}, 2024.

\bibitem{chen_compositional_2024}
\BIBentryALTinterwordspacing
T.~Chen, Z.~Shen, B.~Li, and X.~Xue, ``\BIBforeignlanguage{en}{Compositional scene modeling with global object-centric representations},'' \emph{\BIBforeignlanguage{en}{Machine Learning}}, vol. 113, no.~6, pp. 3505--3524, Jun. 2024. [Online]. Available: \url{https://doi.org/10.1007/s10994-023-06419-5}
\BIBentrySTDinterwordspacing

\bibitem{webb_systematic_2023}
\BIBentryALTinterwordspacing
T.~Webb, S.~S. Mondal, and J.~D. Cohen, ``\BIBforeignlanguage{en}{Systematic {Visual} {Reasoning} through {Object}-{Centric} {Relational} {Abstraction}},'' \emph{\BIBforeignlanguage{en}{Advances in Neural Information Processing Systems}}, vol.~36, pp. 72\,030--72\,043, Dec. 2023. [Online]. Available: \url{https://proceedings.neurips.cc/paper_files/paper/2023/hash/e3cdc587873dd1d00ac78f0c1f9aa60c-Abstract-Conference.html}
\BIBentrySTDinterwordspacing

\bibitem{zeghidour_wavesplit_2021}
N.~Zeghidour and D.~Grangier, ``Wavesplit: {End}-to-{End} {Speech} {Separation} by {Speaker} {Clustering},'' \emph{IEEE/ACM Transactions on Audio, Speech, and Language Processing}, vol.~29, pp. 2840--2849, 2021, conference Name: IEEE/ACM Transactions on Audio, Speech, and Language Processing.

\bibitem{StollerED18-0}
\BIBentryALTinterwordspacing
D.~Stoller, S.~Ewert, and S.~Dixon, ``Wave-u-net: A multi-scale neural network for end-to-end audio source separation,'' in \emph{Proceedings of the 19th International Society for Music Information Retrieval Conference, ISMIR 2018, Paris, France, September 23-27, 2018}, E.~Gómez, X.~H. 0001, E.~Humphrey, and E.~Benetos, Eds., 2018, pp. 334--340. [Online]. Available: \url{http://ismir2018.ircam.fr/doc/pdfs/205_Paper.pdf}
\BIBentrySTDinterwordspacing

\bibitem{wisdom_whats_2021}
\BIBentryALTinterwordspacing
S.~Wisdom, H.~Erdogan, D.~P.~W. Ellis, R.~Serizel, N.~Turpault, E.~Fonseca, J.~Salamon, P.~Seetharaman, and J.~R. Hershey, ``What’s all the {Fuss} about {Free} {Universal} {Sound} {Separation} {Data}?'' in \emph{{ICASSP} 2021 - 2021 {IEEE} {International} {Conference} on {Acoustics}, {Speech} and {Signal} {Processing} ({ICASSP})}, Jun. 2021, pp. 186--190, iSSN: 2379-190X. [Online]. Available: \url{https://ieeexplore.ieee.org/document/9414774/?arnumber=9414774}
\BIBentrySTDinterwordspacing

\bibitem{petermann_hyperbolic_2023}
\BIBentryALTinterwordspacing
D.~Petermann, G.~Wichern, A.~Subramanian, and J.~L. Roux, ``Hyperbolic {Audio} {Source} {Separation},'' in \emph{{ICASSP} 2023 - 2023 {IEEE} {International} {Conference} on {Acoustics}, {Speech} and {Signal} {Processing} ({ICASSP})}, Jun. 2023, pp. 1--5, iSSN: 2379-190X. [Online]. Available: \url{https://ieeexplore.ieee.org/document/10094943/?arnumber=10094943}
\BIBentrySTDinterwordspacing

\bibitem{liu_separate_2023}
\BIBentryALTinterwordspacing
X.~Liu, Q.~Kong, Y.~Zhao, H.~Liu, Y.~Yuan, Y.~Liu, R.~Xia, Y.~Wang, M.~D. Plumbley, and W.~Wang, ``\BIBforeignlanguage{en}{Separate {Anything} {You} {Describe}},'' Oct. 2023, arXiv:2308.05037 [cs, eess]. [Online]. Available: \url{http://arxiv.org/abs/2308.05037}
\BIBentrySTDinterwordspacing

\bibitem{yuan_compositional_2023}
J.~Yuan, T.~Chen, B.~Li, and X.~Xue, ``Compositional {Scene} {Representation} {Learning} via {Reconstruction}: {A} {Survey},'' \emph{IEEE Transactions on Pattern Analysis and Machine Intelligence}, pp. 1--20, 2023, conference Name: IEEE Transactions on Pattern Analysis and Machine Intelligence.

\bibitem{locatello_object-centric_2020}
\BIBentryALTinterwordspacing
F.~Locatello, D.~Weissenborn, T.~Unterthiner, A.~Mahendran, G.~Heigold, J.~Uszkoreit, A.~Dosovitskiy, and T.~Kipf, ``Object-{Centric} {Learning} with {Slot} {Attention},'' in \emph{Advances in {Neural} {Information} {Processing} {Systems}}, vol.~33.\hskip 1em plus 0.5em minus 0.4em\relax Curran Associates, Inc., 2020, pp. 11\,525--11\,538. [Online]. Available: \url{https://proceedings.neurips.cc/paper/2020/hash/8511df98c02ab60aea1b2356c013bc0f-Abstract.html}
\BIBentrySTDinterwordspacing

\bibitem{carion_end--end_2020}
N.~Carion, F.~Massa, G.~Synnaeve, N.~Usunier, A.~Kirillov, and S.~Zagoruyko, ``\BIBforeignlanguage{en}{End-to-{End} {Object} {Detection} with {Transformers}},'' in \emph{\BIBforeignlanguage{en}{Computer {Vision} – {ECCV} 2020}}, ser. Lecture {Notes} in {Computer} {Science}, A.~Vedaldi, H.~Bischof, T.~Brox, and J.-M. Frahm, Eds.\hskip 1em plus 0.5em minus 0.4em\relax Cham: Springer International Publishing, 2020, pp. 213--229.

\bibitem{seitzerbridging}
M.~Seitzer, M.~Horn, A.~Zadaianchuk, D.~Zietlow, T.~Xiao, C.-J. Simon-Gabriel, T.~He, Z.~Zhang, B.~Sch{\"o}lkopf, T.~Brox \emph{et~al.}, ``Bridging the gap to real-world object-centric learning,'' in \emph{The Eleventh International Conference on Learning Representations}.

\bibitem{singh_simple_2022}
\BIBentryALTinterwordspacing
G.~Singh, Y.-F. Wu, and S.~Ahn, ``\BIBforeignlanguage{en}{Simple {Unsupervised} {Object}-{Centric} {Learning} for {Complex} and {Naturalistic} {Videos}},'' \emph{\BIBforeignlanguage{en}{Advances in Neural Information Processing Systems}}, vol.~35, pp. 18\,181--18\,196, Dec. 2022. [Online]. Available: \url{https://proceedings.neurips.cc/paper_files/paper/2022/hash/735c847a07bf6dd4486ca1ace242a88c-Abstract-Conference.html}
\BIBentrySTDinterwordspacing

\bibitem{brady_provably_2023}
\BIBentryALTinterwordspacing
J.~Brady, R.~S. Zimmermann, Y.~Sharma, B.~Schölkopf, J.~V. Kügelgen, and W.~Brendel, ``\BIBforeignlanguage{en}{Provably {Learning} {Object}-{Centric} {Representations}},'' in \emph{\BIBforeignlanguage{en}{Proceedings of the 40th {International} {Conference} on {Machine} {Learning}}}.\hskip 1em plus 0.5em minus 0.4em\relax PMLR, Jul. 2023, pp. 3038--3062, iSSN: 2640-3498. [Online]. Available: \url{https://proceedings.mlr.press/v202/brady23a.html}
\BIBentrySTDinterwordspacing

\bibitem{engelcke_genesis-v2_2021}
\BIBentryALTinterwordspacing
M.~Engelcke, O.~Parker~Jones, and I.~Posner, ``{GENESIS}-{V2}: {Inferring} {Unordered} {Object} {Representations} without {Iterative} {Refinement},'' in \emph{Advances in {Neural} {Information} {Processing} {Systems}}, vol.~34.\hskip 1em plus 0.5em minus 0.4em\relax Curran Associates, Inc., 2021, pp. 8085--8094. [Online]. Available: \url{https://proceedings.neurips.cc/paper/2021/hash/43ec517d68b6edd3015b3edc9a11367b-Abstract.html}
\BIBentrySTDinterwordspacing

\bibitem{singh2022illiterate}
\BIBentryALTinterwordspacing
G.~Singh, F.~Deng, and S.~Ahn, ``Illiterate {DALL}-e learns to compose,'' in \emph{International Conference on Learning Representations}, 2022. [Online]. Available: \url{https://openreview.net/forum?id=h0OYV0We3oh}
\BIBentrySTDinterwordspacing

\bibitem{reddy_audioslots_2023}
\BIBentryALTinterwordspacing
P.~Reddy, S.~Wisdom, K.~Greff, J.~R. Hershey, and T.~Kipf, ``Audioslots: {A} {Slot}-{Centric} {Generative} {Model} {For} {Audio} {Separation},'' in \emph{2023 {IEEE} {International} {Conference} on {Acoustics}, {Speech}, and {Signal} {Processing} {Workshops} ({ICASSPW})}, Jun. 2023, pp. 1--5. [Online]. Available: \url{https://ieeexplore.ieee.org/abstract/document/10193208}
\BIBentrySTDinterwordspacing

\bibitem{gha_unsupervised_2023}
\BIBentryALTinterwordspacing
J.~Gha, V.~Herrmann, B.~Grewe, J.~Schmidhuber, and A.~Gopalakrishnan, ``Unsupervised {Musical} {Object} {Discovery} from {Audio},'' Nov. 2023, arXiv:2311.07534 [cs, eess]. [Online]. Available: \url{http://arxiv.org/abs/2311.07534}
\BIBentrySTDinterwordspacing

\bibitem{huang_masked_2022}
\BIBentryALTinterwordspacing
P.-Y. Huang, H.~Xu, J.~Li, A.~Baevski, M.~Auli, W.~Galuba, F.~Metze, and C.~Feichtenhofer, ``\BIBforeignlanguage{en}{Masked {Autoencoders} that {Listen}},'' \emph{\BIBforeignlanguage{en}{Advances in Neural Information Processing Systems}}, vol.~35, pp. 28\,708--28\,720, Dec. 2022. [Online]. Available: \url{https://proceedings.neurips.cc/paper_files/paper/2022/hash/b89d5e209990b19e33b418e14f323998-Abstract-Conference.html}
\BIBentrySTDinterwordspacing

\bibitem{wu_inverted-attention_2023}
\BIBentryALTinterwordspacing
Y.-F. Wu, K.~Greff, G.~F. Elsayed, M.~C. Mozer, T.~Kipf, and S.~v. Steenkiste, ``\BIBforeignlanguage{en}{Inverted-{Attention} {Transformers} can {Learn} {Object} {Representations}: {Insights} from {Slot} {Attention}},'' Oct. 2023. [Online]. Available: \url{https://openreview.net/forum?id=m9s6rnYWqm}
\BIBentrySTDinterwordspacing

\bibitem{salamon_scaper_2017}
\BIBentryALTinterwordspacing
J.~Salamon, D.~MacConnell, M.~Cartwright, P.~Li, and J.~P. Bello, ``\BIBforeignlanguage{en}{Scaper: {A} library for soundscape synthesis and augmentation},'' in \emph{\BIBforeignlanguage{en}{2017 {IEEE} {Workshop} on {Applications} of {Signal} {Processing} to {Audio} and {Acoustics} ({WASPAA})}}.\hskip 1em plus 0.5em minus 0.4em\relax New Paltz, NY: IEEE, Oct. 2017, pp. 344--348. [Online]. Available: \url{http://ieeexplore.ieee.org/document/8170052/}
\BIBentrySTDinterwordspacing

\bibitem{Sridhar2023}
S.~Sridhar and M.~Cartwright, ``Multi-label open-set audio classification,'' in \emph{Proceedings of the 8th Detection and Classification of Acoustic Scenes and Events 2023 Workshop (DCASE2023)}, Tampere, Finland, September 2023, pp. 171--175.

\bibitem{akiba_optuna_2019}
\BIBentryALTinterwordspacing
T.~Akiba, S.~Sano, T.~Yanase, T.~Ohta, and M.~Koyama, ``\BIBforeignlanguage{en}{Optuna: {A} {Next}-generation {Hyperparameter} {Optimization} {Framework}},'' in \emph{\BIBforeignlanguage{en}{Proceedings of the 25th {ACM} {SIGKDD} {International} {Conference} on {Knowledge} {Discovery} \& {Data} {Mining}}}.\hskip 1em plus 0.5em minus 0.4em\relax Anchorage AK USA: ACM, Jul. 2019, pp. 2623--2631. [Online]. Available: \url{https://dl.acm.org/doi/10.1145/3292500.3330701}
\BIBentrySTDinterwordspacing

\bibitem{rousseeuw_silhouettes_1987}
\BIBentryALTinterwordspacing
P.~J. Rousseeuw, ``Silhouettes: {A} graphical aid to the interpretation and validation of cluster analysis,'' \emph{Journal of Computational and Applied Mathematics}, vol.~20, pp. 53--65, Nov. 1987. [Online]. Available: \url{https://www.sciencedirect.com/science/article/pii/0377042787901257}
\BIBentrySTDinterwordspacing

\bibitem{wisdom_unsupervised_2020}
\BIBentryALTinterwordspacing
S.~Wisdom, E.~Tzinis, H.~Erdogan, R.~Weiss, K.~Wilson, and J.~Hershey, ``Unsupervised {Sound} {Separation} {Using} {Mixture} {Invariant} {Training},'' in \emph{Advances in {Neural} {Information} {Processing} {Systems}}, vol.~33.\hskip 1em plus 0.5em minus 0.4em\relax Curran Associates, Inc., 2020, pp. 3846--3857. [Online]. Available: \url{https://proceedings.neurips.cc/paper/2020/hash/28538c394c36e4d5ea8ff5ad60562a93-Abstract.html}
\BIBentrySTDinterwordspacing

\end{thebibliography}

\end{document}